\documentclass{article}

\pdfoutput=1

\usepackage{booktabs}
\usepackage{hyphenat}
\usepackage{flushend}
\usepackage{graphicx}
\usepackage{enumitem}
\usepackage{cite}
\usepackage{csquotes}
\usepackage{siunitx}

\usepackage{multirow}
\usepackage{subfig} 
\usepackage{float}  
\usepackage{caption}
\usepackage{makecell}
\usepackage{amsmath}
\usepackage{amstext}
\newtheorem{Define}{Definition}
\usepackage{breqn}
\usepackage{algorithm}
\usepackage[noend]{algpseudocode}
\algnewcommand{\LeftComment}[1]{\Statex \(\triangleright\) #1} 
\algnewcommand\algorithmicinput{\textbf{Input}}
\algnewcommand\algorithmicoutput{\textbf{Output}}
\algnewcommand\Input{\item[\algorithmicinput]}%
\algnewcommand\Output{\item[\algorithmicoutput]}%
\algdef{SE}[VARIABLES]{Variables}{EndVariables}
    {\algorithmicvariables}
    {\algorithmicend\ \algorithmicvariables}
\algnewcommand{\algorithmicvariables}{\textbf{Global variables}}

\usepackage{relsize}
\usepackage{amssymb}

\let\emptyset\varnothing

\usepackage{ifthen}
\usepackage{xcolor}
\title{ShiftsReduce: Minimizing Shifts in Racetrack Memory 4.0}

\author{
  Asif Ali Khan\thanks{Authors are with the Chair for Compiler Construction, TU Dresden, Germany.} \and 
  Fazal Hameed\footnotemark[1] \and 
  Robin Bl{\"a}sing\thanks{Authors are with the Max Planck Institute of Microstructure Physics at Halle, Germany.} \and
  Stuart Parkin\footnotemark[2] \and Jeronimo Castrillon\footnotemark[1] 
}

\begin{document}
\maketitle
%
\begin{abstract}
\emph{Racetrack memories} (RMs) have significantly evolved since their conception in 2008, making them a serious contender in the field of emerging memory technologies.
Despite key technological advancements, the access latency and energy consumption of an RM-based system are still highly influenced by the number of \emph{shift} operations.
These operations are required to move bits to the right positions in the racetracks. 
This paper presents data placement techniques for RMs that maximize the likelihood that consecutive references access nearby memory locations at runtime thereby minimizing the number of shifts.
We present an \emph{integer linear programming} (ILP) formulation for optimal data placement in RMs, and revisit existing offset assignment heuristics, originally proposed for random-access memories.
We introduce a novel heuristic tailored to a realistic RM and combine it with a genetic search to further improve the solution. 
We show a reduction in the number of shifts of up to 52.5\%, outperforming the state of the art by up to 16.1\%.
\end{abstract}

%



\section{Introduction}
\label{sec:intro}
Conventional SRAM/DRAM-based memory systems are unable to conform to the growing demand of low power, low cost and large capacity memories. 
Increase in the memory size is barred by technology scalability as well as leakage and refresh power. 
As a result, multiple non-volatile memories such as \emph{phase change memory} (PCM), \emph{spin transfer torque} (STT-RAM) and
\emph{resistive RAM} (ReRAM) have emerged and attracted considerable attention~\cite{PCM,STTRAM_onur, STTRAM_ours,ReRAM}. 
These memory technologies offer power, bandwidth and scalability features amenable 
to processor scaling. However, they pose new challenges such as imperfect reliability and higher write latency. 
The relatively new spin-orbitronics based \emph{racetrack memory} (RM) represents a promising option to surmount the aforementioned limitations by
offering ultra-high capacity, energy efficiency, lower per bit cost, higher reliability
and smaller read/write latency~\cite{stuart1.0, stuart4.0}.
Due to these attractive features, RMs have been investigated at all levels in the memory hierarchy. 
Table~\ref{tab:memcomp} provides a comparison of RM with contemporary volatile and non-volatile memories.

The diverse memory landscape has motivated research on hardware and software optimizations for 
more efficient utilization of NVMs in the memory subsystem. 
To avoid the design complexity added by hardware solutions, software-based data placement 
has become an important emerging area for compiler optimization~\cite{dp_survey}. 
Even modern days processors such as intel's Knight Landing Processor offer means for software managed 
on-board memories.
Compiler guided data placement techniques have been proposed at various levels in the memory hierarchy, not only for improving the temporal/spatial locality of the memory objects 
but also the lifetime and high write latency of NVMs~\cite{mac, rthms,semantics_for_dataplacement,hybrid_mem_dataplacement}. 
In the context of \emph{near data processing} (NDP), efficient data placement improves the effectiveness of NDP cores by allowing more accesses to the local memory stack 
and mitigating remote accesses. 

In this paper, we study data placement optimizations for the particular case of racetrack memories. 
While RMs do not suffer from reliability and latency issues, they pose a significantly different challenge. 
From the architectural perspective, RMs store multiple bits ---1 to 100--- per access point in the form of \emph{magnetic domains} in a tape-like structure, referred to as \emph{track}. 
Each track is equipped with one or more \emph{magnetic tunnel junction} (MTJ) sensors, referred to as \emph{access ports}, that are used to perform read/write operations. 
While a track could be equipped with multiple access ports, the number of access ports per track are always much smaller than the number of domains. 
In the scope of this paper, we consider the ideal single access port per track for ultra high density of the RM. 
This implies that the desired bits have to be shifted and aligned to the port positions prior to their access. 
The shift operations not only lead to variable access latency but also impact the energy consumption of a system, since the time and the energy required for an access depend on the position of the domain relative to the access port.
We propose a set of techniques that reduce the number of shift operations by placing temporally close accesses at nearby locations inside the RM.


\begin{table}[tb]
\centering

\caption{Comparison of RM with other memory technologies~\cite{Survey_Memories, stuart4.0}}
\centering
\label{tab:memcomp}

\scalebox{0.63}
{%
\begin{tabular}{|c|c|c|c|c|c|c|c|}
\hline
			& SRAM 			& eDRAM	 	& DRAM	& 
STT-RAM		& ReRAM		& PCM		& RaceTrack 4.0\\  \hline
Cell Size ($F^2$) 	& 120-200 		& 30-100	& 4-8	& 6-50	
		& 4-10		& 4-12		& $\leq $ 2\\  \hline
Write Endurance 	& $\geq$ $10^{16}$ 	& $\geq$ $10^{16}$	& 
$\geq$ $10^{16}$	& 4 X 
$10^{12}$		& $10^{11}$	& $10^9$ & $10^{18}$	\\  \hline
Read Time 		& Very Fast 		& Fast			& 
Medium		& Medium	& Medium		& Slow & Fast   \\  \hline
Write Time 		& Very Fast 		& Fast		& Medium	
& Slow	
		& Slow		& Very Slow & Fast\\  \hline
Dynamic Write Energy 	& Low 			& Medium	& Medium	
& High	
		& High		& High		  & Low\\  \hline
Dynamic Read Energy 	& Low 			& Medium	& Medium	
& Low	
		& Low		& Medium	 & Low \\  \hline
Leakage Power 		& High 			& Medium	& Medium	
& Low	
		& Low		& Low		 & Low \\  \hline
Retention Period 	& As long as  	& $30-100$ $\mu$s	& $64-512$ ms &  Variable		& Years		& Years		  & Years\\  
                    & volt applied 	& 	&  &  		& 		& 		  & \\  \hline

\hline
\end{tabular}}
\end{table}
Concretely, we make the following contributions.
\begin{enumerate}
\item An \emph{integer linear programming} (ILP) formulation of the data placement problem for RMs. 

\item A thorough analysis of existing offset assignment heuristics, originally proposed for data placement in DSP stack frames, for data placement in RM.

\item \emph{ShiftsReduce}, a heuristic that computes memory offsets by exploiting the temporal locality of accesses.

\item An improvement in the state-of-the-art RM-placement heuristic~\cite{chen2016} to judiciously decide the next memory offset in case of multiple contenders. 

\item A final refinement step based on a genetic algorithm to further improve the results. 
\end{enumerate}

We compare our approach with existing solutions on the OffsetStone benchmarks~\cite{OffsetStone}.
ShiftsReduce diminishes the number of shifts by 28.8\% 
which is 4.4\% and 6.6\% better than the best performing heuristics~\cite{OffsetStone} and~\cite{chen2016} respectively.

The rest of the paper is organized as follows.
Section~\ref{sec:background} explains the recently proposed RM 4.0, provides motivation for this work and reviews existing data placement heuristics.
Our ILP formulation and the ShiftsReduce heuristic are described in 
Section~\ref{sec:ilp} and Section~\ref{sec:heuristics} respectively. 
Benchmarks description, evaluation results and analysis are presented in Section~\ref{sec:results}.
Section~\ref{sec:related_work} discusses state-of-the-art and Section~\ref{sec:conclusion} concludes the paper.

\section{Background and motivation}
\label{sec:background}
This section provides background on the working principle of RMs, current 
architectural sketches and further motivates the data placement problem (both for 
RAMs and RMs). 

\subsection{Racetrack memory}
\label{subsec:RM}
\begin{figure}[bth]
\centering
\includegraphics[scale=0.45]{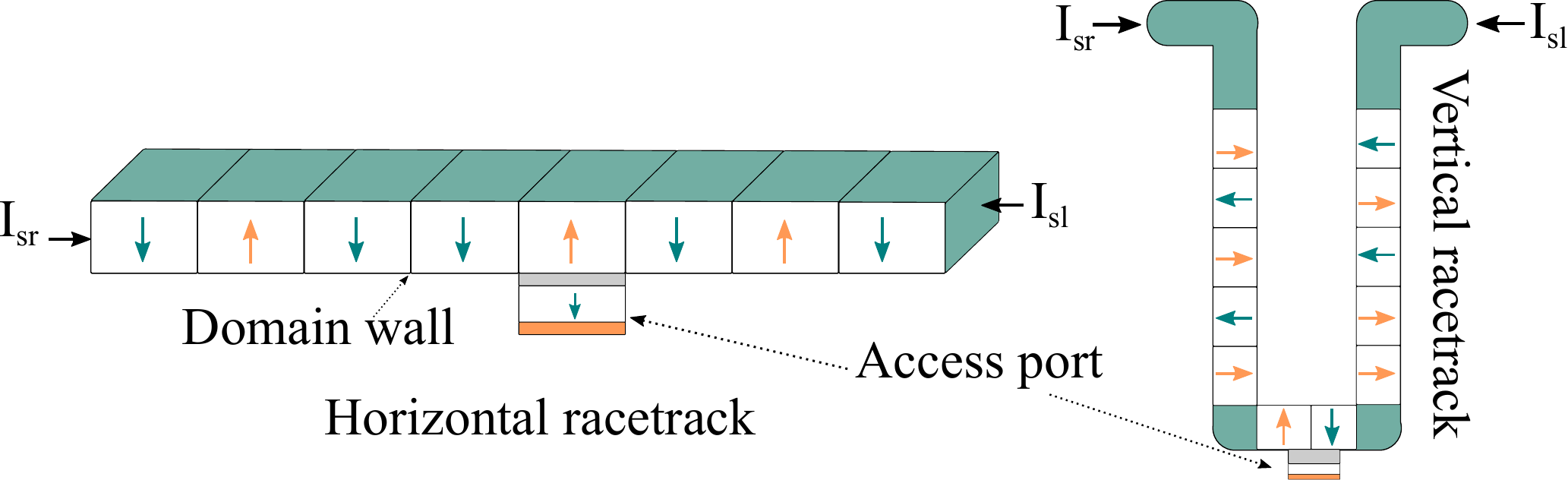}
\caption{Racetrack horizontal and vertical placements (I$_{sl}$ and I$_{sr}$ represent left and right shift currents respectively)}
\label{fig:RM}
\end{figure}

Memory devices have evolved over the last decades from hard disk drives to novel spin-orbitronics based memories. 
The latter uses spin-polarized currents to manipulate the state of the memory. 
The domain walls (DWs) in RMs are moved into a third dimension by an electrical current~\cite{Parkin_patent_2004, stuart1.0}. %
The racetracks can be placed vertically (3D) or horizontally (2D) on the surface of a silicon wafer as shown in Fig.~\ref{fig:RM}.
This allows for higher density but is constrained by crucial design factors such as the shift speed, the DW-to-DW distance and insensitivity to external influences 
such as magnetic fields. 

In earlier RM versions, DWs were driven by a current through a magnetic layer which attained a DW velocity of about 100 ms$^{-1}$~\cite{Hayashi_RM_07}. 
The discovery of even higher DW velocities in structures where the magnetic film was grown on top of a heavy metal allowed to increase the DW velocity to about 300 ms$^{-1}$~\cite{Miron_2011}. 
The driving mechanism is based on spin-orbit effects in the heavy metal which lead to spin currents injected into the magnetic layer~\cite{Ryu_2013}. 
However, a major drawback of these designs was that the magnetic film was very sensitive to external magnetic fields.
Furthermore, they exhibited fringing fields which did not allow to pack DWs closely to each other.

The most recent RM 4.0 resolved these issues by adding an additional magnetic layer on top, which fully compensates the magnetic moment of the bottom layer.
As a consequence, the magnetic layer does not exhibit fringing fields and is insensitive to external magnetic fields. 
In addition, due to the exchange coupling of the two magnetic layers, the DWs velocity can reach up to 1000 ms$^{-1}$~\cite{Yan_2015, stuart4.0}.

\begin{figure}[bth]
\centering
\includegraphics[scale=0.6]{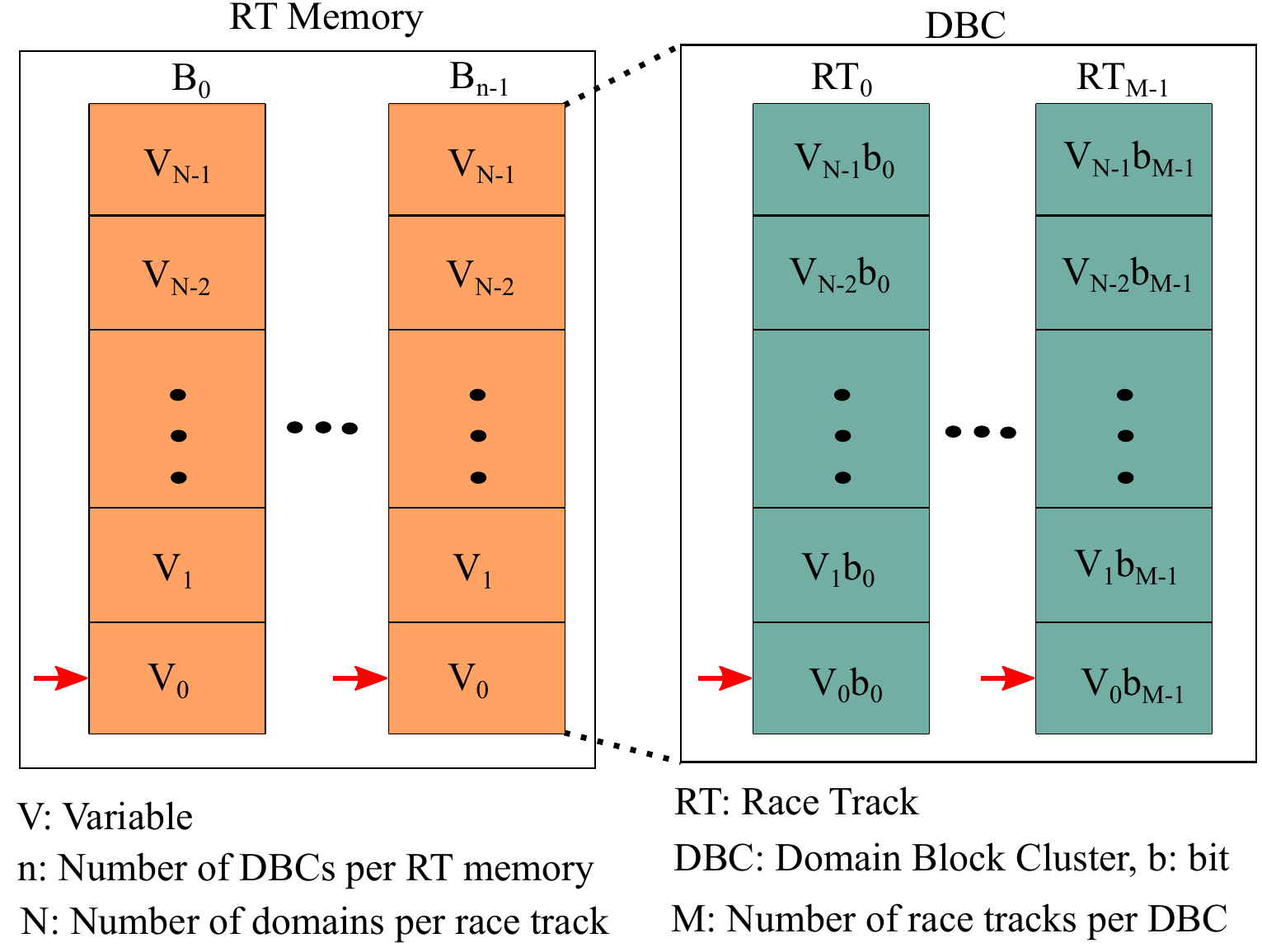}
\caption{Racetrack memory architecture~\cite{tapcache}}
\label{fig:dbc}
\end{figure}

\subsubsection{Memory architecture}
\label{sec:mem_arch}
Fig.~\ref{fig:dbc} shows a widespread architectural sketch of an RM based on~\cite{tapcache}.
In this architecture an RM is divided into multiple \emph{Domain Block Clusters} (\emph{DBCs}), 
each of which contains $M$ tracks with $N$ DWs each.
Each domain wall stores a single bit, and we assume that each \emph{M-bit} variable is 
distributed across $M$ tracks of a DBC.
Accessing a bit from a track requires shifting and aligning the corresponding domain to the track's port position.
We further assume that the domains of all tracks in a particular DBC move in a lock step fashion so that all $M$ bits of a variable are aligned to the port position at the same time for 
simultaneous access. 
We consider a single port per track because adding more ports increases the area.
This is due to the use of additional transistors, decoders, sense amplifiers and output drivers.
As shown in Fig.~\ref{fig:dbc}, each DBC can store a maximum of $N$ variables.

Under the above assumptions, the shift cost to access a particular variable may vary from 0 to $N - 1$.
It is worth to mention that worst case shifts can consume more than 50\% of the RM energy~\cite{zhang2015} and prolong access latency by 26x compared to SRAM~\cite{tapcache}. The architectural simulator, RTSim~\cite{rtsim}, can be used to analyze the shifts' impact on the RM performance and energy consumption, and explore its design space by varying the above mentioned design parameters. 

\begin{figure}[tbh]
        \centering 
     \subfloat[Program variables and access sequence\label{subfig:access_seq}]{%
       \includegraphics[width=0.4\textwidth]{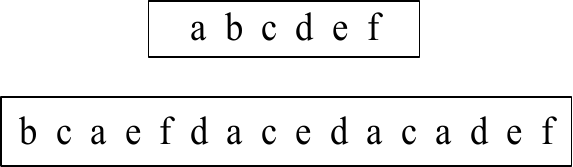}
     }
     \\
     \subfloat[Data placements\label{subfig:placements}]{%
       \includegraphics[width=0.6\textwidth]{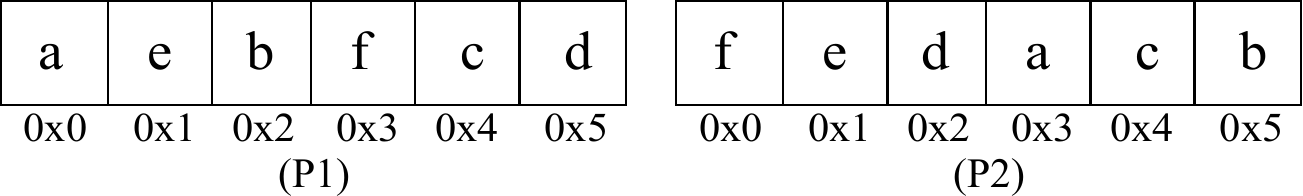}
     }
     \caption{Motivation example}
     \label{fig:motivation}
\end{figure}

\subsection{Motivation example}
\label{subsec:motivation}
To illustrate the problem of data placement consider the set of data items and their access order from Fig.~\ref{subfig:access_seq}. 
We refer to the set of program data items as the set of \emph{program variables} (${\mathcal{V}}$) and  
the set of their access order as \emph{access sequence} ($S$), where $S_i \in {\mathcal{V}} \text{ } \forall{i \in \{0,1,\dots,|S|-1\}}$, for any given source code. 
Note that data items can refer to actual variables placed on a function stack or to accesses to 
fields of a structure or elements of an array.
We assume two different, a naive (P1) and a more carefully chosen (P2), memory placements of the program variables as shown in Fig.~\ref{subfig:placements}. 
\begin{figure}[tbh]
\centering
\includegraphics[width=0.65\textwidth]{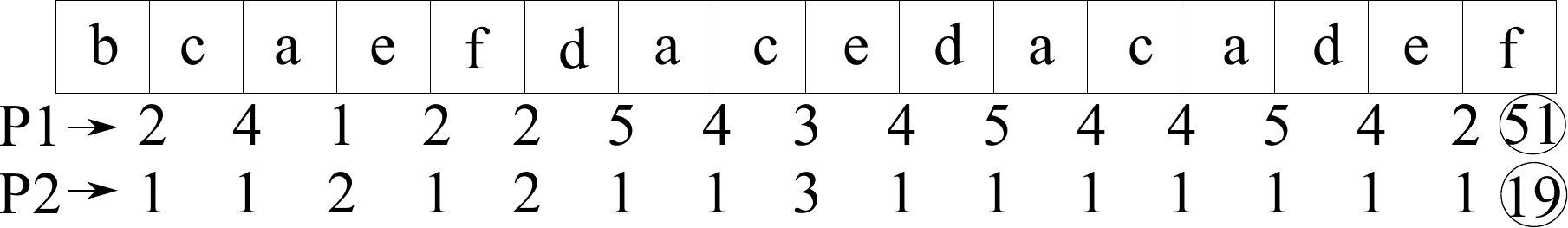}
\caption{Number of shifts in placements P1 and P2 from Fig~\ref{subfig:placements} (encircled numbers show the total shift cost)}
\label{subfig:shifts_cost}
\end{figure}

The number of shifts for the two different placements, P1 and P2 in Fig.~\ref{subfig:placements}, 
are shown in Fig.~\ref{subfig:shifts_cost}. 
The shift cost between any two successive accesses in the access sequence is equivalent 
to the absolute difference of their memory offsets (e.g, $|2-4|$ for b,c in P1). 
The naive data placement P1 incurs $51$ shifts in accessing the entire access sequence, 
while P2 incurs only $19$, i.e., $2.6\times$ better. 
This leads to an improvement in both latency and energy consumption for the simple illustrative
example. 

\begin{figure}[bth]
\centering
\includegraphics[scale=0.7]{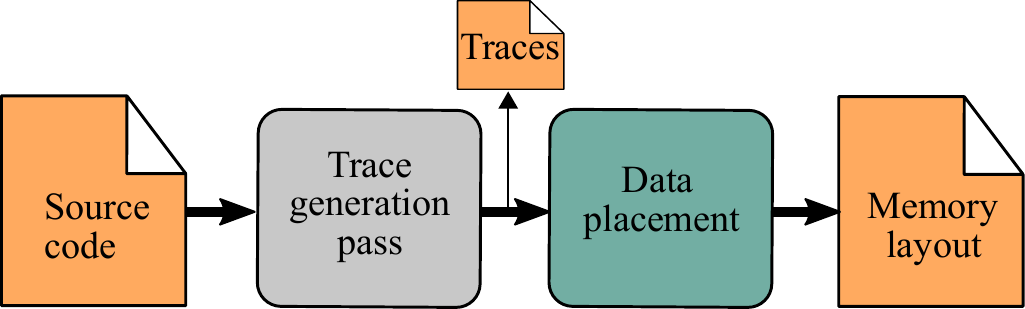}
\caption{Data placement in RMs}
\label{fig:trace_flow}
\end{figure}

\subsection{Problem definition}
\label{subsec:problem_def}
Fig.~\ref{fig:trace_flow} shows a conceptual flow of the data placement problem in RMs. 
The access sequence corresponds to memory traces which can be obtained with standard techniques.
They can be obtained via profiling and tracing, e.g., using Pin~\cite{pin}, inferred from 
static analysis, e.g., for \emph{Static Control Parts} using polyhedral analysis, or with 
a hybrid of both as in~\cite{trace_gen}. 
In this paper we assume the traces are given and focus on the data placement step to produce 
the memory layout. 
We investigate a number of exact/inexact solutions that intelligently decide \emph{memory offsets of the program variables referred to as memory layout} based on the access sequence.  
The memory for which the layout is generated could either be a scratchpad memory, a software 
managed flat memory similar to the on-board memory in intel's Knight Landing Processor or the memory stack exposed to an NDP core. 

The shift cost of an access sequence depends on the memory offsets of the data items. 
We assume that each data item is stored in a single memory offset of the RM (cf. Section~\ref{sec:mem_arch}).
We denote the memory offset of a data item $u \in {\mathcal{V}}$ as $\beta(u)$. 
The shift cost between two data items $u$ and $v$ is then: 
\begin{equation}
\Delta(u, v) = |\beta(u) - \beta(v)| \quad \forall{u,v \in {\mathcal{V}}}
\label{eq:distance}
\end{equation}
The total shift cost ($C$) of an access sequence ($S$) is computed by accumulating the shift costs of successive accesses: 
\begin{equation}
C = \left( \sum_{i=0}^{|S|-2} \Delta(S_i, S_{i+1}) \right)
\label{eq:cost_across_seq}
\end{equation}
The data placement problem for RMs can be then defined as:
\begin{Define}
\label{def:shiftcost}
Given a set of variables ${\mathcal{V}} = \{v_0, v_1,,\dots,v_{n-1}\}$ and an access 
sequence $S = (S_0, S_1,\dots,S_{m-1}),\ S_i \in {\mathcal{V}}$, 
find a data placement $\beta$ for ${\mathcal{V}}$ such that the total cost $C$ is minimized. 
\end{Define}

\subsection{State-of-the-art data placement solutions}
\label{subsec:soa}
\begin{figure}[tbh]
\centering
\includegraphics[scale=0.6]{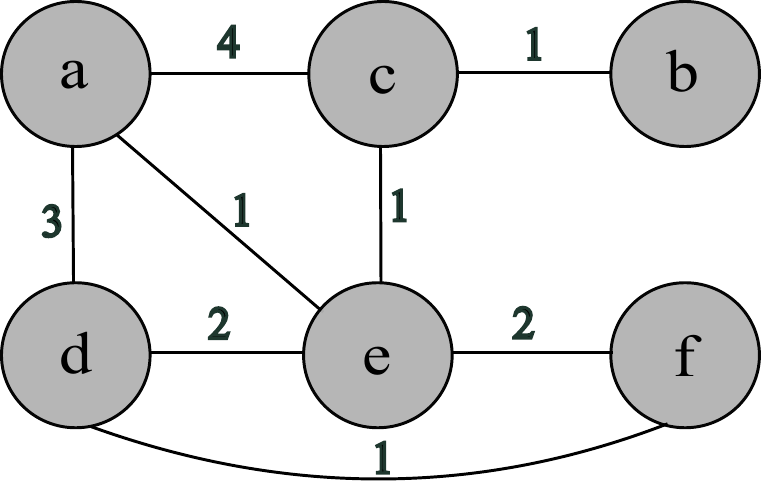}
\caption{Access graph for the access sequence in Fig.~\ref{subfig:access_seq}}
\label{fig:access_graph_2}
\end{figure}

The data placement problem in RMs is similar to the classical single offset assignment (SOA) problem in DSP's stack frames~\cite{OffsetStone, liao, bartley, INC}. 
The heuristics proposed for SOA assign offsets to stack variables; aiming at maximizing the likelihood that 
two consecutive references at runtime will be to the same or 
adjacent stack locations.
Most SOA heuristics work on an \emph{access graph} and formulate the problem as maximum weighted Hamiltonian path (MWHP) or maximum weight path covering (MWPC). 
An access graph $G = (V, E)$ represents an access sequence where $V$ is the set of vertices corresponding to program 
variables (${\mathcal{V}}$). 
An edge $ e = \{u, v\} \in E$ has weight $w_{uv}$ if variables $u, v \in {\mathcal{V}}$ are accessed consecutively $w_{uv}$ times in $S$. 
The assignment is then constructed by solving the MWHP/MWPC problem.
The access graph for the access sequence in Fig.~\ref{subfig:access_seq} is shown in Fig.~\ref{fig:access_graph_2}. 

The SOA cost for two consecutive accesses is \emph{binary}. 
That is, if the next access cannot be reached within the 
auto-increment/decrement range, an extra instruction is needed to modify the address 
register (cost of $1$). 
The cost is $0$ otherwise. 
In contrast, the shift cost in RM is a natural number. 
For RM-placement, the SOA heuristics must be revisited since they only consider 
edge weights of successive elements in $S$.
This may produce better results on small access sequences due to the limited number of vertices and smaller end-to-end distance in $S$, but might not perform well 
on longer access sequences.
In this paper, we extend the SOA heuristics to account for the more general cost function. 
\medskip

Chen et al. recently proposed a group-based heuristic for data placement in RMs~\cite{chen2016}. 
Based on an access graph it assigns offsets to vertices by moving 
them to a group $g$. The position of a data item within a group indicates its memory offset. 
The first vertex added to the group has the maximum \emph{vertex-weight} in the access graph where vertex-weight is the sum of all edge weights that connect a vertex to other vertices in $G$. 
The remaining elements are iteratively added to the group, based on their \emph{vertex-to-group weights} (maximum first). 
The vertex-to-group weight 
of a vertex $u$ is the sum of all edge weights that connect $u$ to the 
vertices in $g$.

We argue that intelligent tie-breaking for equal vertex-to-group weights deserves investigation. Further the static assignment of highest weight vertex to offset 0 seems restrictive. 
Defining positions relative to other vertices provides more flexibility to navigate the solution space. 

\section{Optimal data placement: ILP formulation}
\label{sec:ilp}
This section presents an ILP formulation for the data placement problem in RM. 
Consider the access graph $G$ and the access sequence $S$ to variables $ v \in {\mathcal{V}}$, 
the edge weight $w_{v_iv_j}$ between variables $v_i, v_j$ can be computed as: 
\begin{equation}
    w_{v_iv_j}= 
\begin{cases}
    \sum_{x=0}^{m-2} \Upsilon_{ix} \cdot \Upsilon_{j,x+1} + \Upsilon_{jx} \cdot \Upsilon_{i,x+1}, & i \neq j \\
    0,              & i = j  
\end{cases}
\label{eq:edgeweight}
\end{equation}
with $i,j \in \{0,1,..,n-1\}, n = |{\mathcal{V}}|, m = |S|$ and $\Upsilon$ defined as:
\begin{equation}
    \Upsilon_{ix}= 
\begin{cases}
    1,& \text{if } S_x = v_i \\
    0,& \text{otherwise}  
\end{cases}
\end{equation}
To model unique variable offsets we introduce binary variables ($\Theta_{io}$):
\begin{equation}
    \Theta_{io}= 
\begin{cases}
    1, & \textrm{if} \; v_i \; \text{has memory offset} \; o, \; \forall{i,o \in \{0,1,..,n-1\}}  \\
    0,              & \text{otherwise}  
\end{cases}
\end{equation}
The memory offset of $v_i$ is then computed as:
\begin{equation}
\beta(v_i) = \sum_{o=0}^{n-1} \Theta_{io} \cdot o \quad 
\label{eq:loc}
\end{equation}
Since edges in the access graph embodies the access sequence information, we use them to compute the total shift cost as: 
\begin{equation}
C = \left( \sum_{i=0}^{n-1} \sum_{j=i+1}^{n-2}  w_{v_iv_j} \cdot \Delta(v_i, v_j) \right)
\label{eq:cost}
\end{equation}

The cost function in Equation~\ref{eq:cost} is not inherently linear due to the absolute function in $\Delta(v_i, v_j)$ (cf. Equation~\ref{eq:distance}). 
Therefore, we generate new products and perform subsequent linearization. 
We introduce two integer variables $(p_{ij}, q_{ij}) \in \mathbb{Z}$ to rewrite $|\beta(v_i) - \beta(v_j)|$ as: 
\begin{equation} 
\Delta(v_i, v_j) = p_{ij} + q_{ij} \quad \forall{i,j \in \{0,1,..,n-1\}}
\label{eq:absolute_rep}
\end{equation}
such that
\begin{equation}\tag{C1}	  
\beta(v_i) - \beta(v_j) + p_{ij} - q_{ij} = 0
\end{equation}
\begin{equation}\tag{C2}	  
p_{ij} \cdot q_{ij} = 0
\end{equation}

The second non-linear constraint (C2) implies that one of the two integer variables must be $0$.
To linearize it, we use two binary variables $a_{ij}, b_{ij}$ and a set of constraints:
\begin{equation} \tag{C3}
a_{ij} \le p_{ij} \le a_{ij} \cdot n
\end{equation}
\begin{equation} \tag{C4}
b_{ij} \le q_{ij} \le b_{ij} \cdot n
\end{equation}
\begin{equation} \tag{C5}
0 \le a_{ij} + b_{ij} \le 1
\end{equation}

C5 guarantees that the value of both binary variables $a_{ij}$ and $b_{ij}$ can not be $1$ simultaneously for a given pair $i, j$. 
This, in combination with C3-C4, sets one of the two integer variables to $0$. 
We introduce the following constraint to enforce that the offsets assigned to data items are unique:
\begin{equation} \tag{C6}
p_{ij} + q_{ij} \ge 1
\end{equation}
It ensures uniqueness because the left hand side of the constraint is the difference of the two memory locations (cf. Eq.~\ref{eq:absolute_rep}).

Finally, the linear objective function is:
\begin{equation}
C = \text{min}\left( \sum_{i=0}^{n-1} \sum_{j=i+1}^{n-2}  w_{v_iv_j} \cdot (p_{ij} + q_{ij}) \right)
\end{equation}
The following two constraints are added to ensure that offsets are within range. 
\begin{equation}\tag{C7}
0 \le \beta_{i} \le n-1 
\end{equation}
\begin{equation}\tag{C8}
\sum_{i=0}^{i=n-1}\beta(v_i) = \frac{n\cdot (n-1) }{2} 
\end{equation}

\section{Approximate data placement}
\label{sec:heuristics}
In this section we describe our proposed heuristic and use the 
insights of our heuristic to extend the heuristic by Chen~\cite{chen2016}. 

\subsection{The ShiftsReduce heuristic}
\label{sec:shiftsred}
\begin{figure}[tbh]
\centering
\includegraphics[scale=0.65]{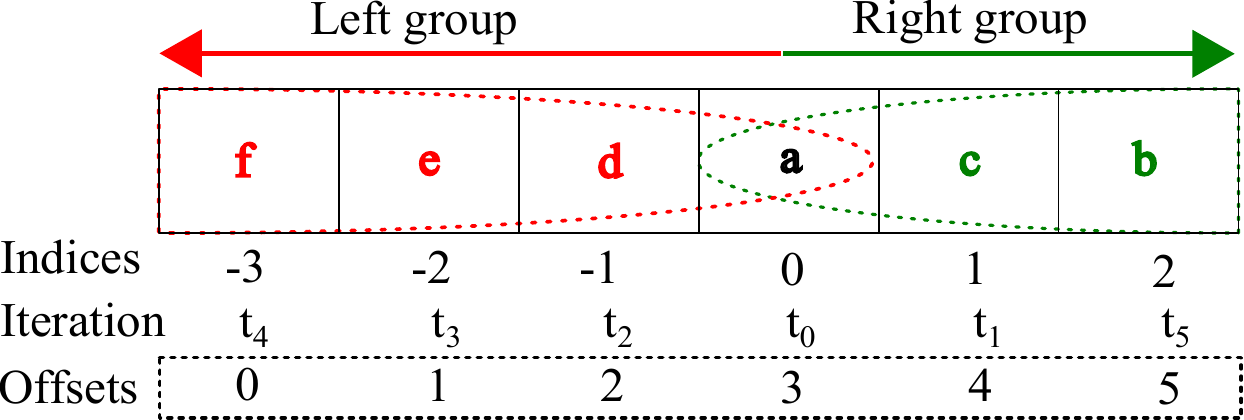}
\caption{Grouping in ShiftsReduce }
\label{fig:ours}
\end{figure}

ShiftsReduce is a group-based heuristic that effectively exploits the locality of accesses in the access sequence and assigns offsets accordingly.
The algorithm starts with the maximum weight vertex in the access graph $G=(V,E)$ and iteratively assigns offsets to the remaining vertices by considering their vertex-to-group weights. 
Recall from Section~\ref{subsec:soa} that the weight of a vertex indicates the count of successive accesses of a vertex with other vertices in $S$, i.e., $w_v=\sum_{u:\{u,v\}\in E}w_{uv}$. 
Note that the maximum weight vertex may not necessarily be the vertex with the highest access frequency, considering repeated accesses of the same vertex. 
\begin{Define}
\label{def:vert_group_weight}
The vertex-to-group weight $\alpha(v,g)$ of a vertex $v \in {\mathcal{V}}$ is defined as the sum of all edge weights that connect $v$ to other vertices in $g$, i.e., 
$\alpha(v, g) = \sum_{u \in g:\{u,v\}\in E} w_{uv}$. 
\end{Define}
ShiftsReduce maintains two groups referred to as left-group $g_l$ (highlighted in red in Fig.~\ref{fig:ours}) and 
right-group $g_r$ (highlighted in green). 
Both $g_l$ and $g_r$ are lists that store the already computed vertices in $V$. 
The heuristic assigns offsets to vertices based on their global and local adjacencies. 
The global adjacency of a vertex $v \in V$ is defined as its vertex-to-group weight with the global group, i.e., $\alpha(v,g_l \cup g_r)$\footnote{We abuse notation, using set operations ($\cup, \setminus$) on lists for better readability.}  
while the local adjacency is the vertex-to-group weight with either of the sub-groups, i.e., $g_l$ or $g_r$.
  
\begin{algorithm}[h]
\caption{ShiftsReduce Heuristic}
\begin{algorithmic}[1]
\Input: Access graph $G = (V, E)$ and a DBC with minimum $n$ empty locations
\Output: Final data placement $\beta$
\State \Comment $v_n$ = fixed element in $g_l$, $v_m$ = fixed element in $g_r$
\State \Comment $v_q$ = last element in $g_l$, $v_p$ = last element in $g_r$
\State $\beta \gets \emptyset, v_{\text{max}} \gets \text{argmax}_{v\in V}w_v$\label{SR:GroupStart}
\State $g_r$.append$(v_{\text{max}})$, $g_l$.append$(v_{\text{max}})$, $V \gets V \setminus \{v_{\text{max}}\}$ \label{SR:index0} 
\State $v^* \gets \text{argmax}_{v\in V}\alpha \text{(}v, g_r\text{)}$\label{SR:V1Compare}
\State  $g_r$.append$(v^*)$, $V \gets V \setminus \{v^*\},v_p \gets v^*$  \label{SR:V1Assign} 
\State $v^* \gets \text{argmax}_{v\in V}\alpha \text{(}v , g_r\setminus \{v^*\}\text{)}$\label{SR:V2Compare}
\State $g_l$.prepend$(v^*)$, $V \gets V \setminus \{v^*\}, v_q \gets v^*$  \label{SR:V2Assign} 
\State $v_n \gets v_{\text{max}}, v_m \gets v_{\text{max}}$ 
\While {$V$ is not empty}\label{SR:LoopStart}
\State $v^* \gets \text{argmax}_{v\in V}\alpha \text{(}v , g_r \cup g_l\text{)}$\label{SR:V1Compare}
  \If {$\alpha (v^*, g_l) > \alpha (v^*, g_r)$}\label{SR:LeftStart} 
  \State $g_l$.prepend$(v^*)$
  \State $\text{(}v_q, v_n\text{)} \gets \textsc{Tie-break(}  v^* ,v_q, v_n, g_l\text{)}$ \label{SR:LeftEnd}
  \ElsIf {$\alpha (v^*, g_l) < \alpha (v^*, g_r)$}\label{SR:RightStart}
  \State $g_r$.append$(v^*)$\State $ \text{(}v_p, v_m\text{)} \gets \textsc{Tie-break(}  v^*, v_p, v_m, g_r\text{)}$\label{SR:RightEnd} 
  \Else\Comment{inter-group tie}\label{SR:Tie} 
      \If {$w_{v^*v_q} > w_{v^*v_p}$} \label{SR:TieBreakIfStart}
      \State $g_l$.prepend$(v^*)$\State $ \text{(}v_q, v_n\text{)} \gets \textsc{Tie-break(}  v^* , v_q, v_n, g_l\text{)}$ \label{SR:TieBreakIfEnd}
      \Else \label{SR:TieBreakElseStart} 
      \State $g_r$.append$(v^*)$\State $ \text{(}v_p, v_m\text{)} \gets \textsc{Tie-break(}  v^*, v_p, v_m, g_r\text{)}$ \label{SR:TieBreakElseEnd}
      \EndIf
  \EndIf
\State $V \gets V \setminus \{v^*\}$
\EndWhile\label{SR:LoopEnd}
\State $\textsc{Assign-offsets$($}\beta , g_l.$append($ g_r.$tail()))
\end{algorithmic}
\label{algo:shiftsreduce}
\end{algorithm}

Pseudocode for the ShiftsReduce heuristic is shown in Algorithm~\ref{algo:shiftsreduce}. 
The sub-groups $g_l$ and $g_r$ initially start at index $0$, the only shared index between $g_l$ and $g_r$, and expand in opposite directions as new elements are added to them. 
We represent this with negative and positive indices respectively as shown in Fig.~\ref{fig:ours}. 
The algorithm selects the maximum weight vertex ($v_{\text{max}}$) and places it at index $0$ in both sub-groups (cf. lines~\ref{SR:GroupStart}-\ref{SR:index0}).

The algorithm then determines two more nodes and add them to the right (cf. line~\ref{SR:V1Assign}) and left (cf. line~\ref{SR:V2Assign}) groups respectively. 
These two nodes correspond to the nodes with the highest vertex-to-group weight ($\alpha$),
which boils down to the maximum edge weight to $v_{\text{max}}$. 
Lines~\ref{SR:LoopStart}-\ref{SR:LoopEnd} iteratively select the next group element based on its global adjacency (maximum first) and add it to $g_l$ or $g_r$ based on its local adjacency. 
If the local adjacency of a vertex with the left group is greater than that of the right group, it is added to left group (cf. lines~\ref{SR:LeftStart}-\ref{SR:LeftEnd}). 
Otherwise, the vertex is added to the right group (cf. lines~\ref{SR:RightStart}-\ref{SR:RightEnd}).

The algorithm prudently breaks both inter-group and intra-group tie situations.
In an inter-group tie situation (cf. line~\ref{SR:Tie}), when the vertex-to-group weight of the selected vertex is equal with both sub-groups, the algorithm compares the edge weight of the selected vertex $v^*$ with the 
last vertices of both groups ($v_p$ in $g_r$ and $v_q$ in $g_l$)
and favors the maximum edge weight (cf. lines~\ref{SR:TieBreakIfStart}-\ref{SR:TieBreakElseEnd}).

To resolve intra-group ties, we introduce the \textsc{Tie-break} function.
The intra-group tie arises when $v_s$ and $v_k$ have equal vertex-to-group-weights with $g$ (cf. line~\ref{TBF:TBStart} in $\textsc{Tie-break}$).
Since the two vertices have equal adjacency with other group elements, they can be placed in any order. 
We specify their order by comparing their edge weights with the fixed vertex ($v_n$ for $g_l$ and $v_m$ for $g_r$) and prioritize the highest edge weight vertex.
The algorithm checks the intra-group tie for every vertex before assigning it to the left-group (cf. line~\ref{SR:LeftEnd}) or right-group (cf. line~\ref{SR:RightEnd}).

We demonstrate ShiftsReduce in Fig.~\ref{fig:ours} for the example in Fig.~\ref{fig:access_graph_2}. 
Vertex $a$ has the highest vertex weight (equal to 4 + 3 + 1 = 8) and is placed at index $0$ in both sub-groups.
Vertices $c$ and $d$ have maximum edge weights with $a$ and are added to the right and left groups respectively (cf. lines~\ref{SR:V1Assign} and~\ref{SR:V2Assign}).
At this point, the two sub-groups contain two elements each. 
The next vertex $e$ is added to $g_l$ because it has higher local adjacency with $g_l$ compared to $g_r$.
In a similar fashion, $b$ and $f$ are added to $g_r$ and $g_l$ respectively.
ShiftsReduce ensures that vertices at far ends of the two groups have least adjacency (i.e., vertex weights) compared to the vertices that are placed in the middle. 
Note that the number of elements in $g_l$ and $g_r$ may not necessarily be equal. 
Finally, offsets are assigned to vertices based on their group positions as highlighted in Fig. \ref{fig:ours}.

Given that we add vertices to two different groups, there are less occurrences of tie compared to algorithms such as Chen's~\cite{chen2016} where vertices are always added to the same group. 
For comparison reasons, we extend Chen's heuristic with tie-breaking in the following section. 
\begin{algorithm}[tbh]
\begin{algorithmic}[1]
\Function{Tie-break}{$v_s, v_k, v_{\text{fix}}, g$}
   	\If {$\alpha (v_s, g\setminus \{v_k\}) = \alpha (v_k, g\setminus \{v_k\})$}\label{TBF:TBStart}
	      \If {$ w_{v_sv_{\text{fix}}} > w_{v_kv_{\text{fix}}}$}\label{TBF:TBIfStart} 
	      \State $v_{\text{fix}} \gets v_s$ \label{TBF:TBIfEnd}
	      \State swap($v_k,v_s$)\Comment{swap positions of $v_k,v_s$}
	      \Else
	      \State $v_{\text{fix}} \gets v_k$ , $v_k \gets v_s$ \label{TBF:TBElse}
	      \EndIf\label{TBF:TBEnd}
	\Else \label{TBF:NEQWeightCompare}
	      \State $v_{\text{fix}} \gets v_k$ , $v_k \gets v_s$ \label{TBF:NEQWeightAssign}
	\EndIf
	
	\Return $(v_k, v_{\text{fix}})$
\EndFunction
\Procedure{Assign-offsets}{$\beta, g$}
\For{$i \gets 0$ to $n-1$}
\State $var \gets \text{variable represented by vertex $g_i$}$
\State $\beta = \beta \cup \{(var, i)\}$
\EndFor
\EndProcedure
\label{algo:TBFunction}
\end{algorithmic}
\end{algorithm}

\subsection{The Chen-TB heuristic}
\label{subsec:chen_tb}

\begin{algorithm}[h]
\caption{Chen-TB Heuristic}
\begin{algorithmic}[1]
\Input: Access graph $G = (V, E)$ and a DBC with minimum $n$ empty locations
\Output: Final data placement $\beta$
\State \Comment $v_m:$ fixed element in $g$, $v_p:$ last element in $g$
\State $\beta \gets \emptyset$, $v^0 \gets \text{argmax}_{v\in V}w_v$ \label{TB:GroupStart}
\State $g$.append$(v^0), V \gets V \setminus \{v^0\}$ 
\State $v^1 \gets \text{argmax}_{v\in V}\alpha \text{(}v , g\text{)}$ \label{TB:V1Compare}
\State $g$.append$(v^1), V \gets V \setminus \{v^1\}$\label{TB:V1Assign}   
\State $v^2 \gets \text{argmax}_{v\in V}\alpha \text{(}v , g\text{)}$ \label{TB:V2Compare}
\State $g$.append$(v^2), V \gets V \setminus \{v^2\}$\label{TB:V2Assign}   
  \If {$ w_{v^0v^2} > w_{v^1v^2}$}\label{TB:012Start}
    \State $v_m \gets v^0, $ swap($v^0, v^1$)\label{TB:012True}
  \Else
    \State $v_m \gets v^1$
  \EndIf\label{TB:GroupEnd}
  
  \While {$V$ is not empty}\label{TB:LoopStart}
    \State $v^* \gets \text{argmax}_{v\in V}\alpha \textit{($v$, $g$)}$\label{SR:V1Compare}
    \State $v_p \gets g.$last()$,g$.append$(v^*)$
    \State $\text{(}v_p, v_m\text{)} \gets \textsc{Tie-break(} v^*, v_p, v_m, g\text{)}$
    \State $V \gets V \setminus \{v^*\}$
  \EndWhile\label{TB:LoopEnd}
\State $\textsc{Assign-offsets(}\beta , g)$
\end{algorithmic}
\label{algo:chen_tb}
\end{algorithm}

\emph{Chen-TB} is a heuristic that extends Chen's heuristic with the \textsc{Tie-break}
strategy introduced for ShiftsReduce. 
As shown in Algorithm~\ref{algo:chen_tb}, Chen-TB initially adds three vertices to the group in lines~\ref{TB:GroupStart}-\ref{TB:GroupEnd}.
In contrast to Chen, we intelligently swap the order of the first two group elements 
by inspecting their edge weights with the third group element.  
Subsequently, lines~\ref{TB:LoopStart}-\ref{TB:LoopEnd} iteratively decide the position of the new group elements until $V$ is empty.
\begin{figure}[tbh]
\centering
\includegraphics[scale=0.37]{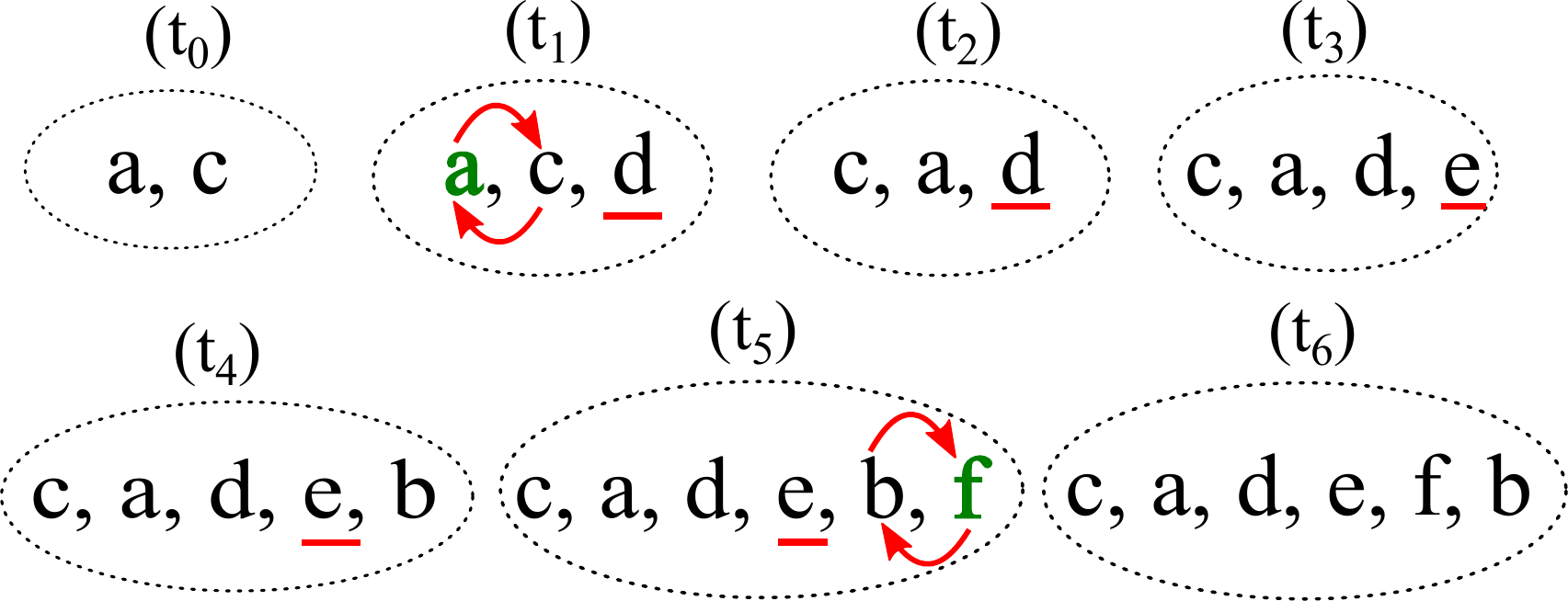}
\caption{Chen-TB heuristic. The fixed element is underlined. The green element has higher edge weight with the fixed element and is moved closer to it. (t$_i$ shows the iteration)}
\label{fig:tb}
\end{figure}

The step-wise addition of vertices to the group is demonstrated in Fig. \ref{fig:tb}.
Initially, the algorithm inspects three vertices from $V$ referred to as $v^0$, $v^1$, and $v^2$.
In line~\ref{TB:GroupStart}, $v^0 = a$ because $a$ has the largest vertex weight ($w_a = 8$).
Next, $v^1 = c$ because $c$ has the maximum edge weight ($w_{ac} = 4$) with $a$ (cf. line~\ref{TB:V1Compare}). 
Similarly, $v^2 = d$ because it has the maximum vertex-to-group weight (which is 3) with $a \cup c$ (cf. line~\ref{TB:V2Compare}).
Since the edge weight between $a$ and $d$ (i.e., $w_{ad}$ = 3)
is higher than the edge weight between $c$ and $d$ (i.e., $w_{cd}$ = 0), 
we swap the positions of $a$ and $c$ in the group (cf. lines~\ref{TB:012Start}-\ref{TB:012True}).
At this point, the group elements are $c, a, d$.
The position of $a$ is fixed while $d$ is the last group element.
The next selected vertex is $e$ due to its highest vertex-to-group weight with $g$.
In this case, the vertex-to-group weight of $d$ and $e$ is compared with $c \cup a$ (cf. line~\ref{TBF:TBStart} in \textsc{Tie-break}).
Since $d$ has higher vertex-to-group weight, $e$ becomes the last element while the position of $d$ is fixed (cf. line~\ref{TBF:NEQWeightAssign} in \textsc{Tie-break}). 
Following the same argument, the next selected element $f$ becomes the last element while the position of $e$ is fixed.
The next selected vertex $b$ and the last element $f$ have equal vertex-to-group-weights i.e. $3$ with the fixed elements $c, a, d , e$. 
Chen-TB prioritizes $f$ over $b$ because it has the higher edge weight with the last fixed element $e$. 

\begin{figure}[tbh]
\centering
\includegraphics[scale=0.63]{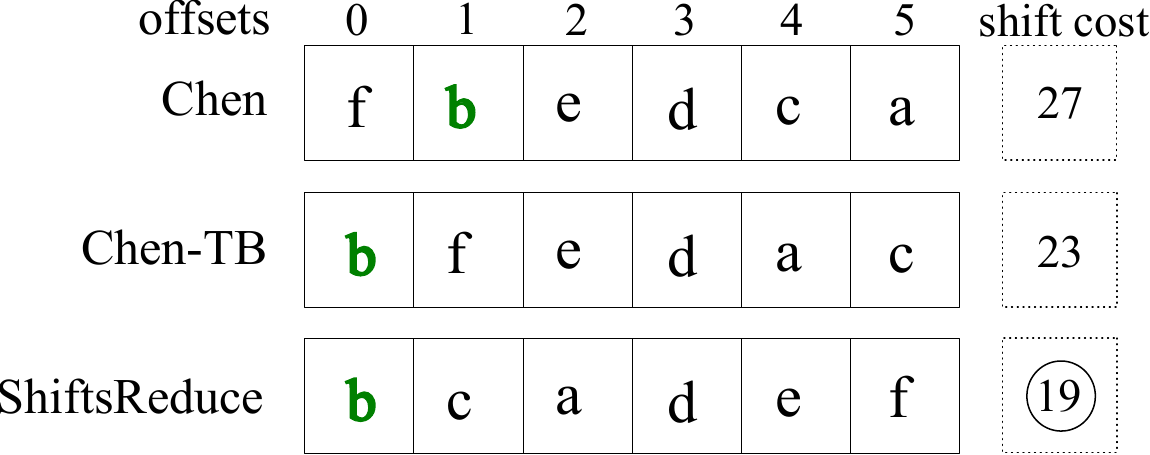}
\caption{Final data placements and costs of Chen, Chen-TB and ShiftsReduce. Initial port 
position marked in green}
\label{fig:all_placements}
\end{figure}

The final data placements of Chen, Chen-TB and ShiftsReduce are presented in Fig.~\ref{fig:all_placements}. 
For the access sequence in Fig.~\ref{fig:access_graph_2}, Chen-TB reduces the number of shifts to 23 compared to 27 by Chen, as shown in Fig.~\ref{fig:all_placements}.
ShiftsReduce further diminishes the shift cost to 19. 
Note that the placement decided by ShiftsReduce is the optimal placement shown in Fig.~\ref{subfig:placements}. 
We assume 3 or more vertices in the access graph for our heuristics because the number of shifts for two vertices, in either order, remain unchanged.  

\section{Results and discussion}
\label{sec:results}
This section provides evaluation and analysis of the proposed solutions on real-world application benchmarks. 
It presents a detailed qualitative and quantitative comparison with state-of-the-art techniques.
Further, it brings a thorough analysis of SOA solutions for RMs. 

\subsection{Experimental setup}
\label{subsec:setup}
We perform all experiments on a Linux Ubuntu (16.04) system with Intel core i7-4790 (3.8~GHz)
processor, $32$~GB memory, g++ v5.4.0 with $-O3$ optimization level. 
We implement our ILP model using the python interface of the Gurobi optimizer, with Gurobi 8.0.1~\cite{gurobi}.

As benchmark we use OffsetStone~\cite{OffsetStone}, which contains more than 3000 realistic 
sequences obtained from complex real-world applications (control-dominated as well as 
signal, image and video processing).
Each application consists of a set of program variables and one or more access sequences. 
The number of program variables per sequence varies from 1 to 1336 while the length of the access sequences lies in the range of 0 and 3640. 
We evaluate and compare the performance of the following algorithms. 
\begin{enumerate}
\item \emph{Order of first use (OFU):} A trivial placement for comparison purposes in which variables are placed in the order they are used.  
\item \emph{Offset assignment heuristics:} For thorough comparison we use Bartley~\cite{bartley},
Liao~\cite{liao}, SOA-TB~\cite{soa_tb}, INC~\cite{INC},  INC-TB~\cite{OffsetStone} and 
the genetic algorithm (GA-SOA) in~\cite{leupers_GA}.
\item \emph{Chen/Chen-TB:} The RM data placement heuristic presented in~\cite{chen2016} and our
extended version (cf. Algorithm~\ref{algo:chen_tb}). 
\item \emph{ShiftsReduce} (cf. Algorithm~\ref{algo:shiftsreduce}).
\item \emph{GA-Ours:} Our modified genetic algorithm for RM data placement described in~\ref{subsec:ga}.
\item \emph{ILP} (cf. Section~\ref{sec:ilp}). 
\end{enumerate}

\subsection{Revisiting SOA algorithms}
\label{subsec:soa}
\begin{figure}[tbh]
\centering
\includegraphics[scale=0.8]{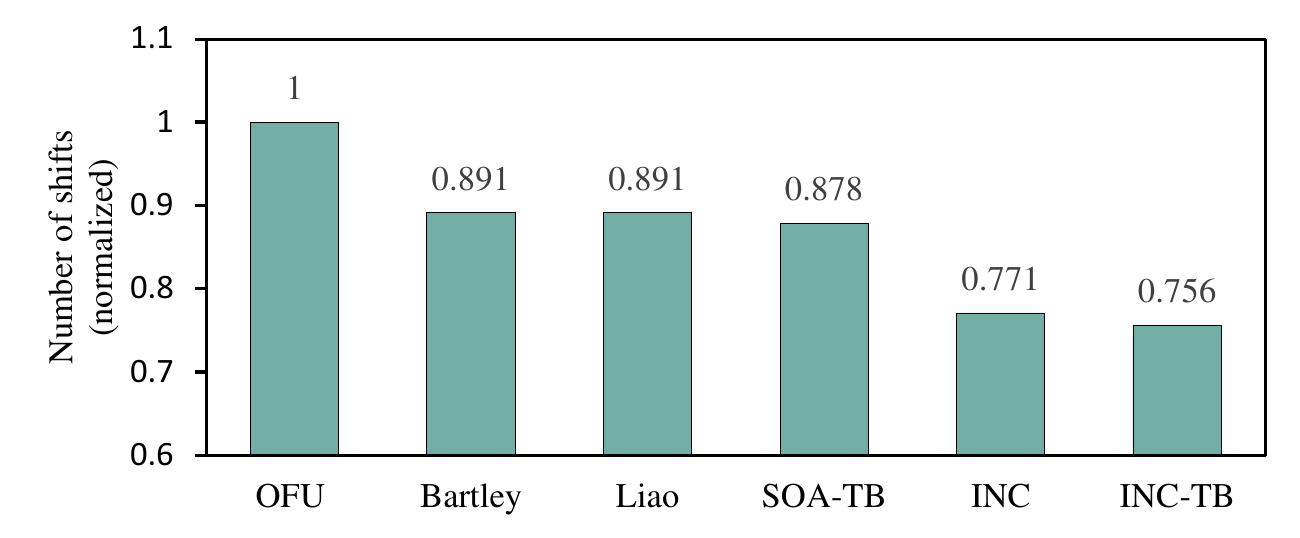}
\caption{Comparison of offset assignment heuristics}
\label{fig:soa}
\end{figure}

We, for the first time, reconsider all well-known offset assignment heuristics. 
The empirical results in Fig.~\ref{fig:soa} show that the SOA heuristics can reduce the shift cost in RM by 24.4\%. 
On average, (Bartley, Liao, SOA-TB, INC and INC-TB) reduce the number of shifts by (10.9\%, 10.9\%, 12.2\%, 22.9\%, 24.4\%) compared to OFU respectively. 
For brevity, we consider only the best performing heuristic i.e., INC-TB for detailed analysis in the following sections. 

\subsection{Analysis of ShiftsReduce}
\label{subsec:SR_analysis}
In the following we analyze our ShiftsReduce heuristic.

\subsubsection{Results overview}
\begin{figure}[tbh]
\includegraphics[scale=0.53]{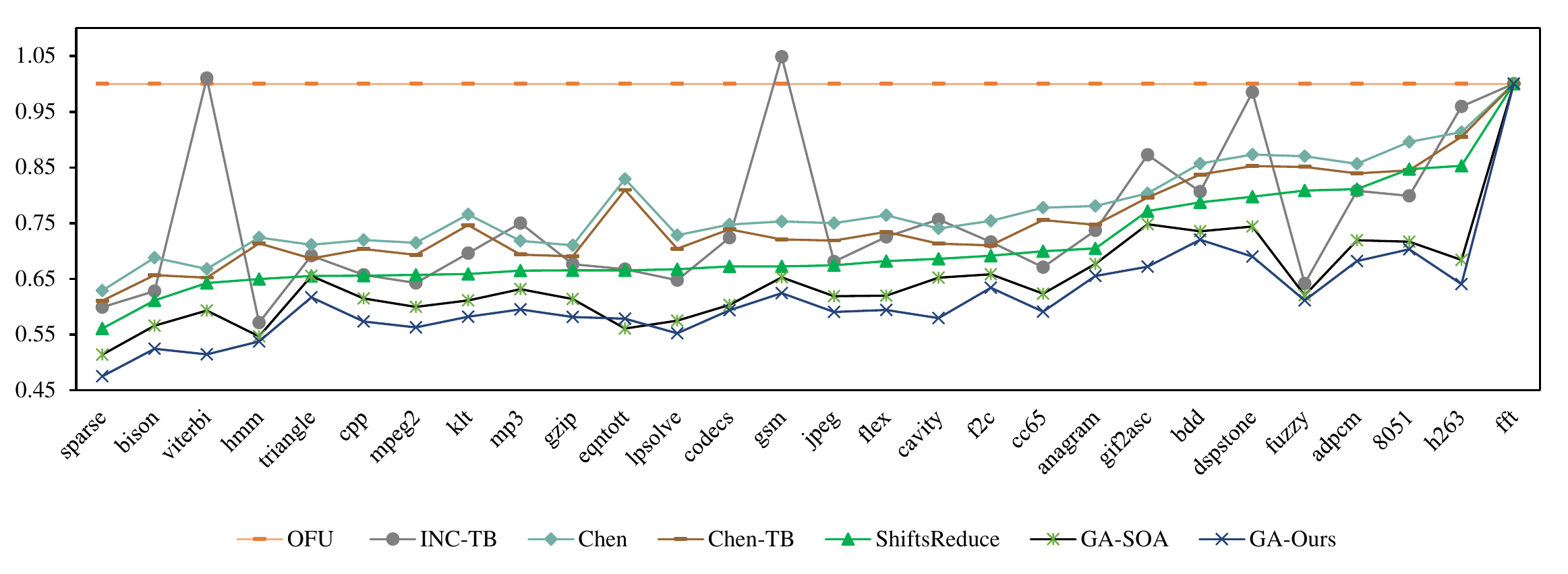}
\caption{Individual benchmark results (sorted in the decreasing order of benefit for ShifsReduce)}
\label{fig:detailed}
\end{figure}

An overview of the results for all heuristics across all benchmarks, normalized to 
the OFU heuristic, is shown in Fig.~\ref{fig:detailed}.
As illustrated, ShiftsReduce yields considerably better performance on most benchmarks. 
It outperforms Chen's heuristic on all benchmarks and INC-TB on 22 out of 28. 
The results indicate that INC-TB underperforms on benchmarks such as \emph{mp3}, \emph{viterbi}, \emph{gif2asc},\emph{dspstone}, and \emph{h263}.
On average, ShiftsReduce curtails the number of shifts by 28.8\% which is 4.4\% and 6.6\% better compared to INC-TB and Chen respectively.

Closer analysis reveals that Chen significantly reduces the shift cost on benchmarks having longer access sequences. 
This is because it considers the global adjacency of a vertex before offset assignment. 
On the contrary, INC-TB maximizes the local adjacencies and favors benchmarks that consist only of shorter sequences. 
ShiftsReduce combines the benefits of both local and global adjacencies, providing superior results.  
None of the algorithms reduce the number of shifts for \emph{fft}, since in this benchmark each 
variable is accessed only once. 
Therefore, any permutation of the variables placement results in identical performance.

\subsubsection{Impact of access sequence length}
\label{subsec:seqs}
\begin{figure}[tbh]
\includegraphics[scale=0.6]{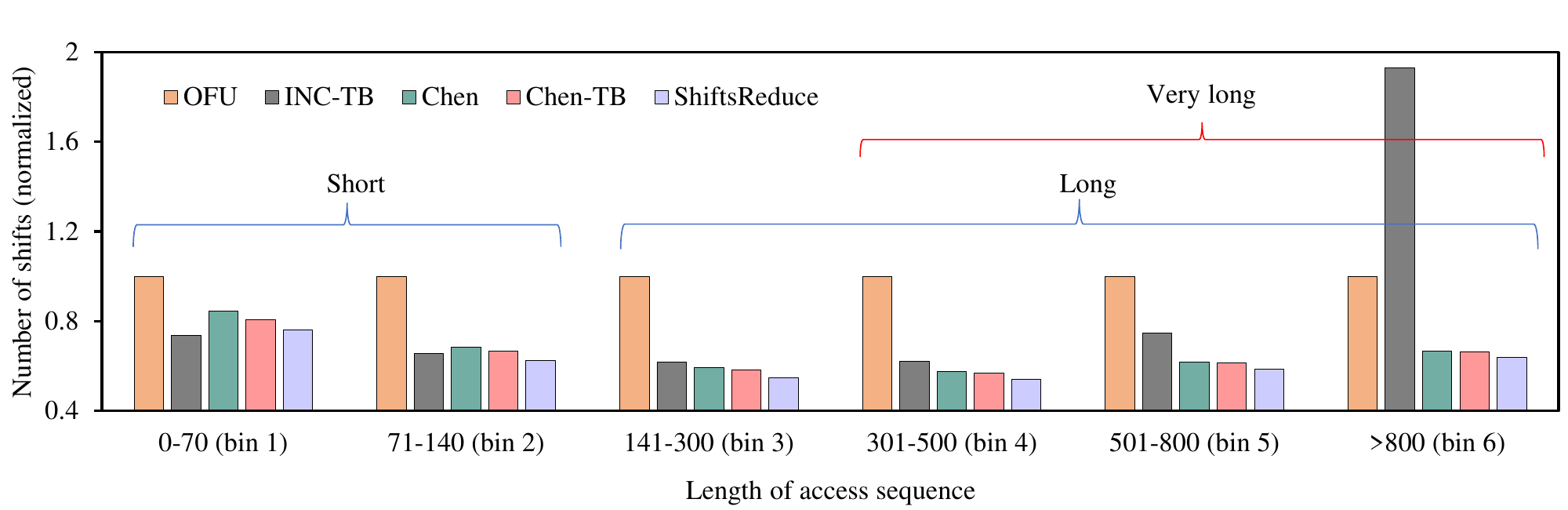}
\caption{Impact of sequence length on heuristic performance}
\label{fig:seq}
\end{figure}
As mentioned above, the length of the access sequence plays a role in the performance
of the different heuristics. 
To further analyze this effect we partition the sequences from all benchmarks in 6 bins based on their lengths. 
The concrete bins and the results are shown in Fig.~\ref{fig:seq}, which reports the average 
number of shifts (smaller is better) relative to OFU.

Several conclusions can be drawn from Fig.~\ref{fig:seq}.
First, INC-TB performs better compared to other heuristics on short sequences. 
For the first bin (0-70), INC-TB reduces the number of shifts by 26.3\% compared to OFU which is 10.9\%, 7.1\% and 2.2\% better than Chen, Chen-TB and 
ShiftsReduce respectively.
Second, the longer the sequence, the better is the reduction compared to OFU. 
Third, the performance of INC-TB aggravates compared to group-based heuristics as the access sequence length increases. 
For bin-5 (501-800), INC-TB reduces the shift cost by 25.2\% compared to OFU while Chen, Chen-TB and 
ShiftsReduce reduces it by 38.3\%, 38.6\% and 41.2\% respectively. 
Beyond 800 (last bin), INC-TB deteriorates performance compared to OFU and even increases the number of shifts by 97.8\%. 
This is due to the fact that INC-TB maximizes memory accesses to consecutive locations (i.e., edge weights) without considering its impact on farther memory accesses (i.e., global adjacency). 
Fourth, Chen performs better compared to INC-TB on long sequences (average 36.6\% for bins 3-6) but falls after it by 6.9\% on short sequences (bins 1-2). 
Fifth, Chen-TB consistently outperforms Chen on all sequence lengths, 
demonstrating the positive impact of the tie-breaking proposed in this paper. 
Finally, the proposed ShiftsReduce heuristic consistently outperforms Chen in all 6 bins. 
This is due to the fact that ShiftsReduce exploit bi-directional group expansion and considers both local and global adjacencies for data placement (cf. Section~\ref{sec:shiftsred}). 
On average, it surpasses (INC-TB, Chen and Chen-TB) by (39.8\%, 3.2\% and 2.8\%) and (0.3\%, 7.3\% and 4.5\%) for long (bins 3-6) and short (bins 1-2) sequences respectively. 

\begin{table}[tbh]
\centering
\caption{Distribution of short, long and very long access sequences in OffsetStone benchmarks}
\centering
\label{tab:benchmarks}
\scalebox{0.9}
{%
\begin{tabular}{|c|c|c|c|c|}
\hline
Category & Benchmarks & Short     & Long  	   & Very Long \\
	 &  	      & Seqs (\%) & Sequences (\%) & Sequences (\%) \\
\hline
\multirow{19}{*}{ $\begin{matrix} \text{category-I} \\ \text{(ShiftsReduce } \\ \text{performs better)}\end{matrix}$ } &	mp3	&	65.1\% &	25.6\% &	9.3\% \\
& veterbi	&	35.0\%   &	40.0\%   &	25.0\% \\
& gif2asc	&	17.7\% &	50.0\%   &	33.3\% \\
& dspstone	&	63.0\% &	29.6\% &	7.4\% \\
& gsm		&	65.1\% &	21.6\% &	13.3\% \\
& cavity	&	20.0\% &	40.0\% &	40.0\% \\
& h263		&	0.0\% &		75.0\% &	25.0\% \\
& codecs	&	59.7\% &	33.3\% &	8.0\% \\
& flex		&	75.8\% &	16.9\% &	7.3\% \\
& sparse	&	69.6\% &	22.8\% &	7.6\% \\
& klt		&	54.5\% &	15.9\% &	29.6\% \\
& triangle	&	75.4\% &	17.2\% &	7.4\% \\
& f2c		&	79.5\% &	15.2\% &	6.3\% \\
& mpeg2		&	50.7\%   &	32.4\% &	16.9\% \\
& bison		&	63.8\% &	26.4\% &	9.8\% \\
& cpp		&	43.7\% &	33.3\% &	13.0\% \\
& gzip		&	50.1\% &	35.2\% &	14.7\% \\
&lpsolve	&	44.6\% &	38.5\% &	16.9\% \\
& jpeg		&	54.5\% &	15.9\% &	29.6\% \\
\hline
\multirow{5}{*}{ $\begin{matrix} \text{category-II} \\ \text{(comparable} \\ \text{performance} \pm2\% \text{)}\end{matrix}$ } 
& bdd		&	85.8\%   &	10.8\%   &	3.4\% \\
& adpcm		&	93.2\% &	3.4\%   &	3.4\% \\
& fft		&	100.0\% &	0.0\% &		0.0\% \\
& anagram	&	100.0\% &	0.0\% &		0.0\% \\
& eqntott	&	100.0\% &	0.0\% &		0.0\% \\
\hline
\multirow{4}{*}{ $\begin{matrix} \text{category-III} \\ \text{(INC performs} \\ \text{ better)} \end{matrix}$ } &	fuzzy	&	100\% &		0.0\% &	0.0\% \\
& hmm		&	79.7\%   &	10.3\%   &	0.0\% \\
& 8051		&	80.0\% &	20.0\%   &	0.0\% \\
& cc65		&	84.6\% &	13.1\% &	2.3\% \\
\hline
\end{tabular}
}
\end{table}
\begin{figure}[tbh]
\centering
\includegraphics[scale=0.8]{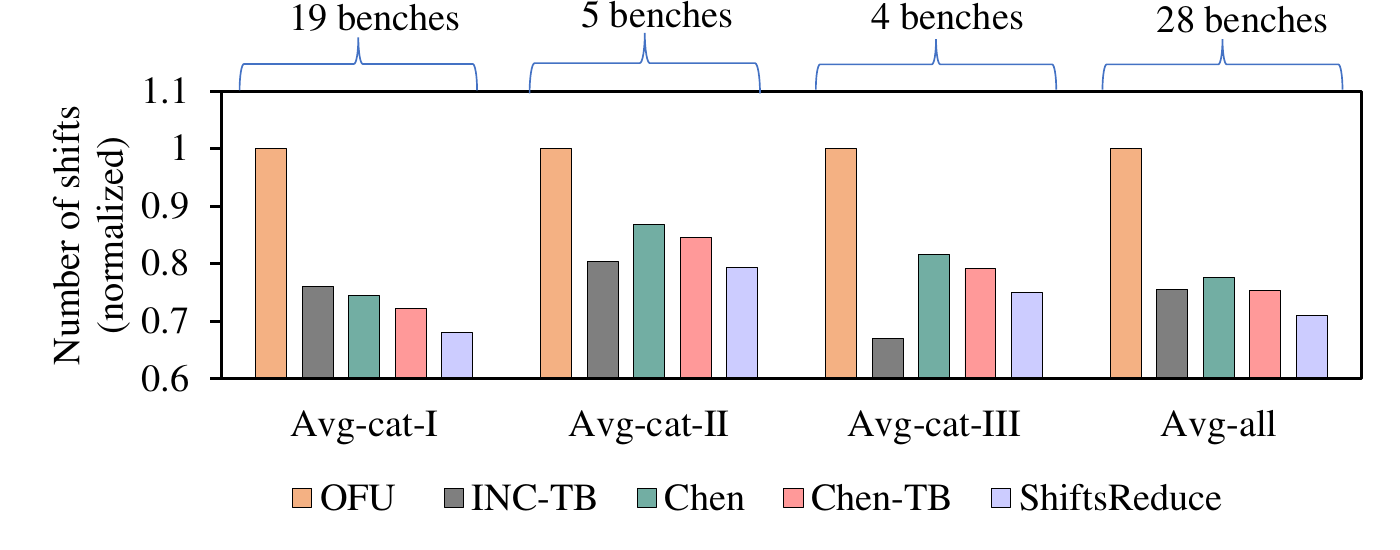}
\caption{Evaluation by benchmark categories}
\label{fig:categories}
\end{figure}

\subsubsection{Category-wise benchmarks evaluation}
\label{subsec:categories}
Based on the above analysis, we classify all benchmarks into 3 categories as shown in Table~\ref{tab:benchmarks}.  
We categorize each access sequence into three ranges i.e., short 
($0-140$), long (greater than $140$) and very-long (greater than $300$). 
The first benchmark category comprises 19 benchmarks; each containing at least 15\% long and 5\% very long access sequences. 
The second and third categories mostly contain short sequences. 

Fig.~\ref{fig:categories} shows that ShiftsReduce provides significant gains on category-I and curtails the number of shifts by 31.9\% (maximum up-to 43.9\%) compared to OFU. 
This is 8.1\% and 6.4\% better compared to INC-TB and Chen respectively.
Similarly, Chen-TB outperforms both Chen and INC-TB by 2.3\% and 4\% respectively.  
INC-TB does not produce good results because the majority of the benchmarks in category-I are dominated by long and/or very long sequences (cf.~Table~\ref{tab:benchmarks} and Section~\ref{subsec:seqs}). 
Category-II comprises 5 benchmarks, mostly dominated by short sequences. 
INC-TB provides higher shift reduction (19.6\%) compared to Chen (13.2\%) and Chen-TB (15.3\%). 
However it exhibits comparable performance with ShiftsReduce (within $\pm2\%$ range). 
On average, ShiftsReduce outperforms INC-TB by 1.1\%. 
INC-TB outperforms ShiftsReduce only on the 4 benchmarks listed in category-III. 

\subsection{GA-SOA vs GA-Ours}
\label{subsec:ga}
Apart from heuristics, \emph{genetic algorithms} (GAs) have also been employed to solve the SOA problem~\cite{leupers_GA}. 
They start with a random population and compute an efficient solution by imitating natural evolution. 
However, GAs always take longer computation times compared to heuristics. 
In order to avoid premature convergence, GAs are often initialized with suboptimal initial solutions. 

This section leverages two genetic algorithms (namely GA-SOA and GA-Ours) for RM data placement. 
We analyze the impact on the results of GA using our solutions compared to solutions obtained with SOA heuristics as initial population. 
Both algorithms use the same parameters as presented in \cite{OffsetStone}. 
The initial populations of GA-SOA and GA-Ours are composed 
of (OFU, Liao~\cite{liao}, INC-TB~\cite{OffsetStone}) and (OFU, Chen-TB, ShiftsReduce) respectively.

Experimental results demonstrate that GA-Ours is superior to GA-SOA in all benchmarks. 
The average reduction in shift cost across all benchmarks (cf. Fig.~\ref{fig:avg_all})
translate to 35.1\% and 38.3\% for GA-SOA and GA-Ours respectively. 

\subsection{ILP results}
\label{subsec:ilp}
\begin{figure}[tbh]
\centering
\includegraphics[scale=0.68]{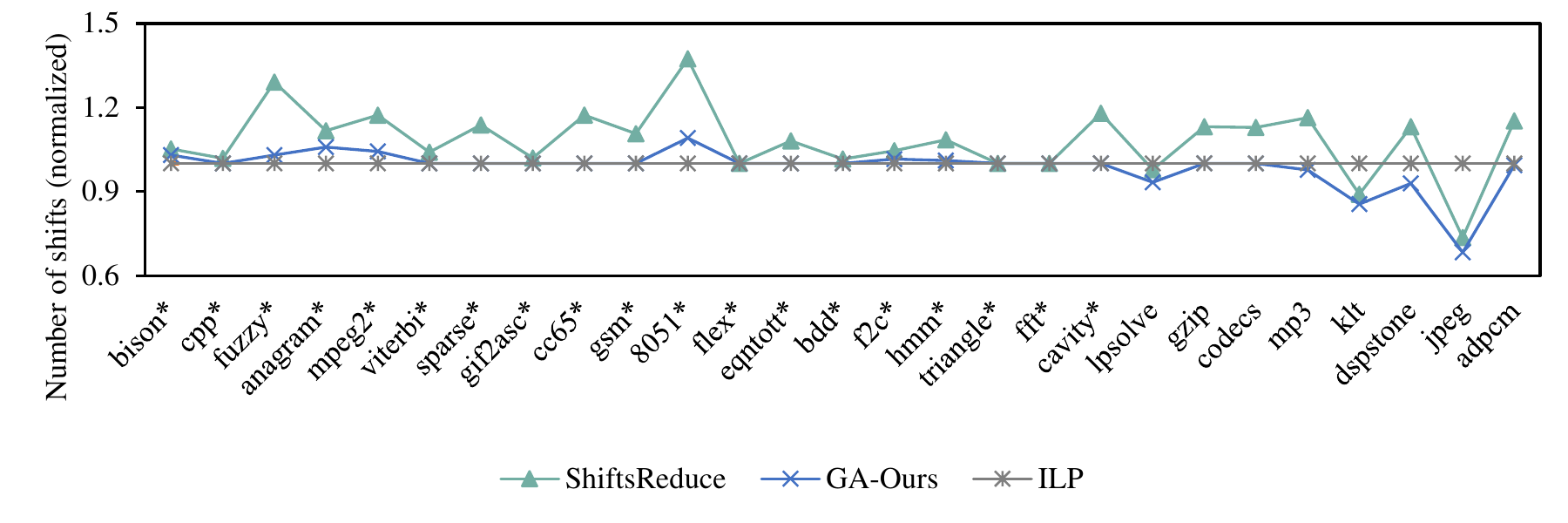}
\caption{Comparison with ILP solution (* mark benchmarks for which an optimal solution was found)}
\label{fig:ilp}
\end{figure}

As expected, the ILP solver could not produce any solution in almost 30\% of the instances when given three hours per instance. 
In the remaining instances, the solver either provides an optimal solution (on shorter sequences) or an intermediate solution. 
We evaluate ShiftsReduce and GA-Ours on those instances where the ILP solver produces results and show the comparison in Fig.~\ref{fig:ilp}. 
On average, the ShiftsReduce results deviate by 8.2\% from the ILP result. GA-Ours bridges this gap and deviate by only 1.3\%.  

\subsection{Summary runtimes and energy analysis}
\label{subsec:total}
\begin{figure}[tbh]
\centering
\includegraphics[scale=0.75]{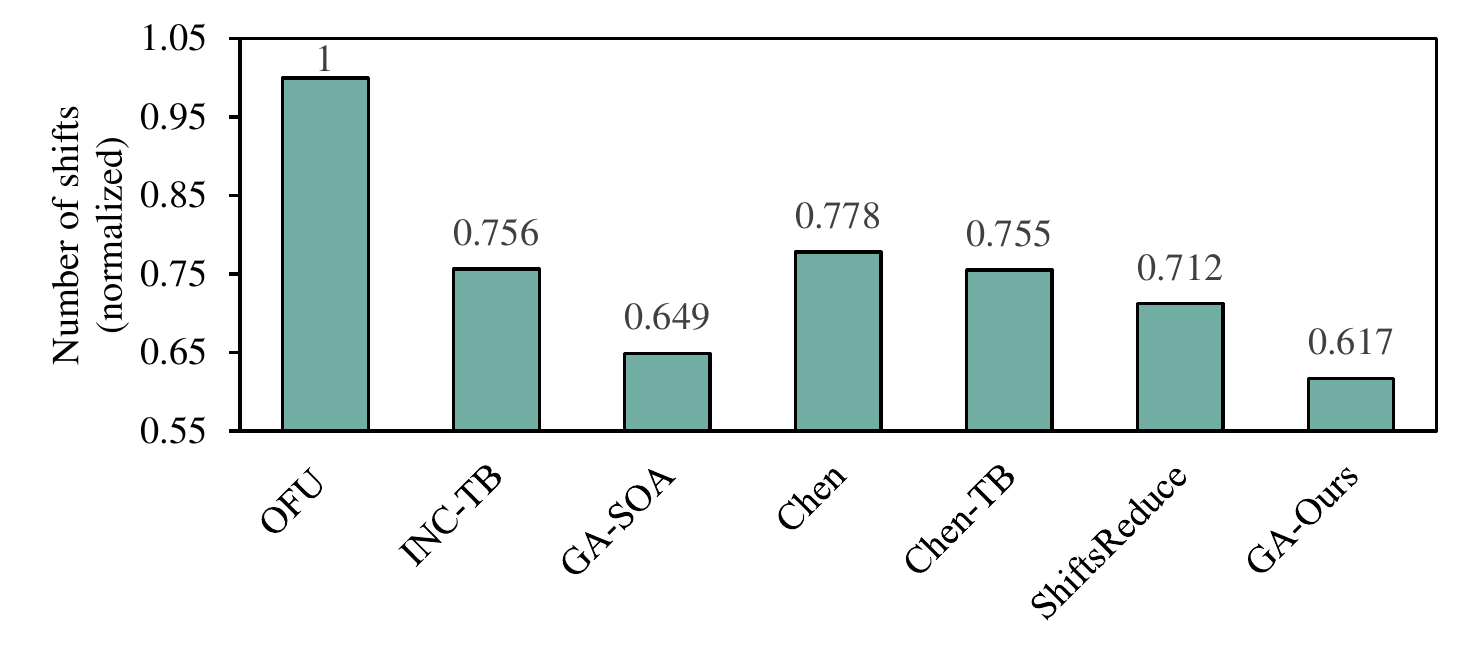}
\caption{Results summary}
\label{fig:avg_all}
\end{figure}

Recall the results overview from Fig.~\ref{fig:avg_all}.
In comparison to OFU, ShiftsReduce and Chen-TB mitigate the number of shifts by 28.8\% and 24.5\% which is (4.4\%, 0.1\%) and (6.6\%, 2.3\%) 
superior than INC-TB and Chen respectively.
Compared to the offset assignment heuristics in Fig.~\ref{fig:soa}, the performance 
improvement of ShiftsReduce and Chen-TB translate to (17.9\%, 17.9\%, 16.6\%, 5.9\%) and (13.6\%, 13.6\%, 12.3\%, 1.6\%) for Bartley, Liao, SOA-TB and INC respectively. 
GA-Ours further reduces the number of shifts in ShiftsReduce by 9.5\%. 
The average runtimes of Chen-TB and ShiftsReduce are 2.99 ms, which is comparable to other 
heuristics, i.e., Bartley (0.23 ms), Liao (0.08 ms), SOA-TB (0.11 ms), INC (2.3 s), INC-TB (2.7 s), GA-SOA (4.96 s), GA-Ours (4.98 s) and Chen (2.98 ms).

Using the latest RM 4.0 prototype device in our in-house physics lab facility,
a current pulse of 1 $ns$, corresponding to a current density of $5 x 10^{11} Amp/m^2$, is applied to the nano-wire to drive the domains.   
Employing a 50 $nm$ wide, 4 $nm$ thick wire, the shift current corresponds to 0.1$mA$. 
With a 5V applied voltage, the power to drive a single domain translates to 0.5 $mW$ ($P = V x I = 5V x 0.1 mA = 0.5 mW$).
Therefore, the energy to shift a single domain amounts to $0.5 pJ$ ($E = P x t = 0.5 mW x 1 ns =  0.5pJ$). 
The RM 4.0 device characteristics indicate the domains in RM 4.0 shifts at a constant velocity without inertial effects.
Therefore, for a 32-bit data item size, the total shift energy amounts to 16$pJ$ without inertia.
The overall shift energy saved by a particular solution is calculated as the total number of shifts for all instances across all benchmark multiplied by per data item shift energy (i.e., 16$pJ$). 
Using RM 4.0, the shift energy reduction for ShiftsReduce relative to OFU translates to 35\%.
In contrast to RM 4.0, the domains in earlier RM prototypes show inertial effects when driven by current.
Considering the inertial effects in earlier RM prototypes, we expect less energy benefits compared to RM 4.0.

\section{Related Work}
\label{sec:related_work}
Conceptually, the racetrack memory is a 1-dimensional version of the classical bubble memory technology of the late 1960s. 
The bubble memory employs a thin film of magnetic material to hold small magnetized areas known as bubbles.
This memory is typically organized as 2-dimensional structure of bubbles composed of major and minor loops~\cite{bubble_ISCA_78}.
The bubble technology could not compete with the Flash RAM due to speed limitations and it vanished entirely by the late 1980s. 
Various data reorganization techniques have been proposed for the bubble memories~\cite{bubble_75, bubble_76, bubble_ISCA_78}.
These techniques alter the relative position of the data items in memory via dynamic reordering so that the more frequently accessed items are close to the access port. 
Since these architectural techniques are blind to exact memory reference patterns of the applications, they might excerbate the total energy consumption. 

Compared to other memory technologies, RMs have the potential to dominate in all performance metrics, for which they have 
received considerable attention as of late.
RMs have been proposed as replacement for all levels in the memory hierarchy for different application scenarios. 
Mao and Wang et al. proposed an RM-based GPU register file to combat the high leakage and scalability problems of conventional SRAM-based 
register files~\cite{gpu_registerfile, gpu_rf}. 
Xu et al. evaluated RM at lower cache levels and reported an energy reduction of 69\% with  comparable performance relative to 
an iso-capacity SRAM~\cite{fusedcache}. 
Sun et al. and Venkatesan  et al. demonstrated RM at last-level cache and reported significant improvements in area (6.4x), energy (1.4x) and
Performance (25\%)~\cite{sun2013, tapcache}. 
Park advocates the usage of RM instead of SSD for graph storage which not only expedites graph processing but also reduces 
energy by up-to 90\%~\cite{ssd}. 
Besides, RMs have been proposed as scratchpad memories~\cite{mao2015}, content addressable memories~\cite{content_addressable} and reconfigurable 
memories~\cite{reconfig_mem}. 

Various architectural techniques have been proposed to hide the RM access latency by pre-shifting the likely accessed DW to the port position~\cite{tapcache}. 
Sun et al. proposed swapping highly accessed DWs with those closer to the access port(s)~\cite{sun2013}. 
Atoofian proposed a predictor-based proactive shifting by exploiting register locality~\cite{predictor_based_preshifting}. 
Likewise, proactive shifting is performed on the data items waiting in the queue~\cite{gpu_rf}.
While these architectural approaches reduce the access latency, they may increase the total number of shifts which exacerbates energy consumption. 

To abate the total number of shifts, techniques such as
data migration~\cite{fusedcache}, data swapping~\cite{sun2013}, data compression~\cite{xu2015}, 
data reorganization for bubble memories~\cite{bubble_75, bubble_76, bubble_ISCA_78}, and efficient 
data placement~\cite {chen2016, mao2015} have been proposed. 
Amongst all, data placement has shown great promise because it effectively reduces the number of shifts with negligible overheads. 

Historically, data placement has been proposed for different memory technologies at different levels in the memory hierarchy. 
It is demonstrated that efficient data placement improves energy consumption and system performance by exploiting temporal/spatial locality of the memory objects ~\cite{cache_con}. 
More recently data placement techniques have been employed in NVM based memory systems in order to improve their performance and lifetimes. 
For instance \cite{hybrid_STTRAM_embedded, mac} employ data placement techniques to hide the higher write latency and hence cache blocks migration overhead in an STT-SRAM hybrid cache. 
Similarly in \cite{rthms, semantics_for_dataplacement, hybrid_mem_dataplacement}, data-placement techniques have been proposed to make efficient 
utilization of the memory systems equipped with multiple memory technologies. 
Likewise, data placement in RMs is proposed for GPU register files~\cite{rm_gpu_rf}, scratchpad memories~\cite{mao2015} 
and stacks~\cite{rm_zpu} in order to reduce the number of shifts. 

In the past, various data placement solutions have been proposed for signal processing in the embedded systems domain (i.e. SOA, cf.~\ref{subsec:soa}). 
These solutions include heuristics~\cite{bartley,liao,soa_tb,INC,OffsetStone}, genetic algorithms~\cite{leupers_GA} and ILP based exact solutions~\cite{tsp_soa, sven_goa_optimal, sven_goa_more_optimal}. 
As discussed in Section~\ref{sec:results} our heuristic builds on top of this previous work, 
providing an improved data allocation.

\section{Conclusions}
\label{sec:conclusion}
This paper presented a set of techniques to minimize the number of shifts in RMs by means of efficient data placement. 
We introduced an ILP model for the data placement problem for an exact solution and heuristic algorithms for efficient solutions.
We show that our heuristic computes near-optimal solutions, at least for small problems, in less than 3 ms.
We revisited well-known offset assignment heuristics for racetrack memories and experimentally showed that they perform better on short access sequences.
In contrast, group-based approaches such as the Chen heuristic exploit global adjacencies and produce better results on longer sequences. 
Our ShiftsReduce heuristic combines the benefits of local and global adjacencies and outperforms all other heuristics, minimizing the number of shifts by up to 40\%. 
ShiftsReduce employs intelligent tie-breaking, a technique that we use to improve the original Chen heuristic.
To further improve the results, we combined ShiftsReduce with a genetic algorithm that improved the results by 9.5\%.  
In future work, we plan to investigate placement decisions together with reordering of accesses from higher abstractions in the compiler, e.g., from a polyhedral model or by exploiting additional semantic information from domain-specific languages. 

\section*{Acknowledgments}
This work was partially funded by the German Research Council (DFG) through the Cluster of Excellence `Center for Advancing Electronics Dresden' (cfaed).
We thank Andr{\'e}s Goens for his useful input in the ILP formulation and Dr. Sven Mallach from Universit{\"a}t zu K{\"o}ln (Cologne) for 
providing the sources of SOA heuristics.

\bibliographystyle{unsrt}
\bibliography{ms}

\end{document}